\documentclass[twocolumn,showpacs,preprintnumbers,amsmath,amssymb]{revtex4}

\usepackage{graphicx}
\usepackage{dcolumn}
\usepackage{bm}

\usepackage{amssymb}
\usepackage{amsmath}

\begin{document}


\title{Generalized thermodynamic constraints \\
on holographic-principle-based cosmological scenarios}

\author{Nobuyoshi {\sc Komatsu}}  \altaffiliation{E-mail: komatsu@se.kanazawa-u.ac.jp} 

\affiliation{Department of Mechanical Systems Engineering, Kanazawa University, 
                          Kakuma-machi, Kanazawa, Ishikawa 920-1192, Japan}

\begin{abstract}

The holographic principle can lead to cosmological scenarios, i.e., holographic equipartition models.
In this model, an extra driving term (corresponding to a time-varying cosmological term) in cosmological equations depends on an associated entropy on the horizon of the universe.
The driving term is expected to be constrained by the second law of thermodynamics, as if the cosmological constant problem could be discussed from a thermodynamics viewpoint.
In the present study, an arbitrary entropy on the horizon, $S_{H}$, is assumed, extending previous analysis based on particular entropies [Phys. Rev. D \textbf{96}, 103507 (2017)].
The arbitrary entropy is applied to the holographic equipartition model, in order to universally examine thermodynamic constraints on the driving term in a flat Friedmann--Robertson--Walker universe at late times.
The second law of thermodynamics for the holographic equipartition model is found to constrain the upper limit of the driving term, even if the arbitrary entropy is assumed.
The upper limit implies that the order of the driving term is likely consistent with the order of the cosmological constant measured by observations.
An approximately equivalent upper limit can be obtained from the positivity of $S_{H}$ in the holographic equipartition model.

\end{abstract}


\maketitle

\section{Introduction} 
\label{Introduction}

Numerous observations imply an accelerated expansion of the late universe \cite{PERL1998_Riess1998,Planck2015,Riess2016}.
The accelerated expansion can be explained by lambda cold dark matter ($\Lambda$CDM) models, which assume an extra driving term related to a cosmological constant $\Lambda$ and dark energy. 
However, $\Lambda$ measured by observations is $120$ orders of magnitude smaller than the theoretical value from quantum field theory, as pointed out repeatedly \cite{Weinberg1989}.
In order to resolve this discrepancy, various models have been proposed \cite{Bamba1}, including $\Lambda (t)$CDM models (i.e., a time-varying $\Lambda (t)$ cosmology) \cite{Freese-Mimoso_2015,Sola_2009-2015,Nojiri2006,Sola_2015L14,LimaSola_2013a,LimaSola_2015_2016,Valent2015,Sola_2013,Sola_2017-2018}, creation of CDM (CCDM) models \cite{Prigogine_1988-1989,Lima1992-1996,LimaOthers2001-2016}, and thermodynamic cosmological scenarios \cite{Sheykhi1,Sadjadi1, Padma2012A,Padma2012,Padma2014-2015,Cai2012-Tu2013,Yuan2013,Tu2013-2015,Krishna20172018,Neto2018a,Hadi-Sheykhia2018,Koma10,Koma11,Sheykhi2,Karami2011-2018,Easson12,Koivisto1Basilakos1-Gohar,Sola_2014a,Gohar_2015a,Gohar_2015b,Koma4,Koma5,Koma6,Koma7,Koma8,Koma9}.
Most thermodynamic scenarios are based on the holographic principle \cite{Hooft-Bousso}, which assumes that the information of the bulk is stored on the horizon.
Among these scenarios, a scenario based on the holographic principle has recently attracted attention.

The  attracted scenario is Padmanabhan's holographic equipartition model \cite{Padma2012A}.
In this model, cosmological equations can be derived from the expansion of cosmic space due to the difference between the degrees of freedom on the surface and in the bulk \cite{Padma2012,Padma2014-2015,Cai2012-Tu2013,Yuan2013,Tu2013-2015,Krishna20172018,Neto2018a,Hadi-Sheykhia2018}.  
However, an extra driving term for the accelerated expansion is not derived from the Bekenstein--Hawking entropy \cite{Bekenstein1Hawking1}, which is usually used for the entropy on the horizon of the universe \cite{Koma10,Koma11}. 
As an alternative to the Bekenstein--Hawking entropy, nonextensive entropies  (such as the Tsallis--Cirto entropy \cite{Tsallis2012} and a modified R\'{e}nyi entropy \cite{Czinner1,Czinner2-3}) and quantum corrections (such as logarithmic corrections \cite{LQG2004_1,LQG2004_2,LQG2004_3} and power-law corrections \cite{Das2008,Radicella2010}) have recently been proposed.
In fact, the modified R\'{e}nyi entropy and the power-law corrected entropy can lead to an extra driving term, as examined by the present author \cite{Koma10,Koma11}.
The driving term corresponding to a time-varying $\Lambda (t)$ is constrained by the second law of thermodynamics \cite{Koma11}. 
This implies that the cosmological constant problem can be discussed from a thermodynamics viewpoint.
However, these results depend on the choice of entropy on the horizon.
Accordingly, not a particular entropy but rather an arbitrary entropy is required for examining thermodynamic constraints on the driving term universally.
The thermodynamic constraint could provide new insights into a discussion of the cosmological constant problem.

In this context, we assume an arbitrary entropy on the horizon, extending the previous works \cite{Koma10,Koma11} (in which particular entropies were used).
In the present study, we apply the arbitrary entropy to the holographic equipartition model, in order to universally examine thermodynamic constraints on an extra driving term in cosmological equations for a flat Friedmann--Robertson--Walker (FRW) universe at late times.
In addition, the driving term is systematically formulated using the Bekenstein--Hawking entropy, which is considered to be the standard scale of the entropy on the horizon.
The formulation should reveal the influence of deviations from the Bekenstein--Hawking entropy on the driving term.
In the present study, we clarify the thermodynamic constraints on the holographic equipartition model and discuss the order of the driving term from a thermodynamics viewpoint.
To this end, both the positivity of entropy and the second law of thermodynamics are examined.
The systematic study of the thermodynamic constraints on this model should help to develop a deeper understanding of cosmological scenarios based on the holographic principle.
Note that we do not discuss the inflation of the early universe because we focus on the late universe.

The remainder of the article is organized as follows.
In Section\ \ref{LCDM models}, $\Lambda (t)$CDM models are briefly reviewed for a typical formulation of cosmological equations.
In Section\ \ref{Bekenstein-Hawking entropy}, the Bekenstein--Hawking entropy is reviewed for the standard scale of the entropy on the horizon of the universe.
In Section\ \ref{Holographic equipartition}, a holographic equipartition model is introduced.
In this section, an acceleration equation that includes an extra driving term is derived, assuming an arbitrary entropy on the horizon.
The driving term is systematically formulated using the Bekenstein--Hawking entropy.
In Section\ \ref{SL}, thermodynamic constraints on the driving term are examined.
The order of the driving term is discussed from both the positivity of entropy and the second law of thermodynamics.
Finally, in Section\ \ref{Conclusions}, the conclusions of the study are presented.

It should be noted that an entropic-force model proposed by Easson \textit{et al.} \cite{Easson12} is one of the cosmological scenarios based on the holographic principle and has been examined from various viewpoints \cite{Koivisto1Basilakos1-Gohar,Sola_2014a,Gohar_2015a,Gohar_2015b,Koma4,Koma5,Koma6,Koma7,Koma8,Koma9}.
The entropic-force model is not discussed in the present study.
(The concept of the entropic-force model is different from the idea that gravity itself is an entropic force \cite{Padma1Verlinde1}.
However, the idea of Verlinde's entropic gravity is applied to the entropic-force model.
For details of entropic gravity, see, e.g., the work of Visser \cite{Visser}.)

\section{$\Lambda(t)$CDM model} 
\label{LCDM models}
Cosmological equations in a holographic equipartition model are expected to be similar to those for $\Lambda (t)$CDM models \cite{Freese-Mimoso_2015,Sola_2009-2015,Nojiri2006,Sola_2015L14,LimaSola_2013a,LimaSola_2015_2016,Valent2015,Sola_2013,Sola_2017-2018}. 
In this section, we briefly review a typical formulation of the cosmological equations for the $\Lambda(t)$CDM model, according to Refs.\ \cite{Koma9,Koma10,Koma11}. 

We consider a flat FRW universe and use the scale factor $a(t)$ at time $t$.
In the $\Lambda(t)$CDM model, the Friedmann equation is given as
\begin{equation}
  \left(  \frac{ \dot{a}(t) }{ a(t) } \right)^2  =  H(t)^2     =  \frac{ 8\pi G }{ 3 } \rho (t) +  f(t)   , 
\label{eq:L(t)_FRW01}
\end{equation}
and the acceleration equation is  
\begin{align}
  \frac{ \ddot{a}(t) }{ a(t) }   &=  \dot{H}(t) + H(t)^{2}                                                                        \notag \\
                                    &=  -  \frac{ 4\pi G }{ 3 } \left ( \rho (t) + \frac{3p(t)}{c^2}  \right ) +  f(t) ,  
\label{eq:L(t)_FRW02}
\end{align}
where the Hubble parameter $H(t)$ is defined by
\begin{equation}
   H(t) \equiv   \frac{ da/dt }{a(t)} =   \frac{ \dot{a}(t) } {a(t)}  .
\label{eq:Hubble}
\end{equation}
Here, $G$, $c$, $\rho(t)$, and $p(t)$ are the gravitational constant, the speed of light, the mass density of cosmological fluids, and the pressure of cosmological fluids, respectively \cite{Koma9}.
Moreover, $f(t)$ is an extra driving term, corresponding to a time-varying cosmological term.

Usually, $f(t)$ is replaced by $\Lambda(t)/3$ for the $\Lambda(t)$CDM model.
When $f(t) =\Lambda/3$, we refer to this model as the $\Lambda$CDM model.
From observations, the order of the density parameter $\Omega_{\Lambda}$ for $\Lambda$ is $1$ \cite{Koma10,Koma11}.
For example, $\Omega_{\Lambda} =0.692$ is from the Planck 2015 results \cite{Planck2015}.
Accordingly, the order of the cosmological constant term, $O(\frac{\Lambda}{3})$, can be written as 
 \begin{equation}
  O \left ( \frac{\Lambda}{3}  \right )   = O \left ( \Omega_{\Lambda}  H_{0}^{2}   \right )  \approx   O  \left ( H_{0}^{2}  \right ) , 
\label{L_order_0}
\end{equation} 
where $\Omega_{\Lambda}$ is defined by $\Lambda /(3 H_{0}^{2})$, and $H_{0}$ is the Hubble parameter at the present time \cite{Koma10,Koma11}.
We later use this to discuss the order of an extra driving term.

Various driving terms for the $\Lambda (t)$CDM model \cite{Freese-Mimoso_2015,Sola_2009-2015,Nojiri2006,Sola_2015L14,LimaSola_2013a,LimaSola_2015_2016,Valent2015,Sola_2013,Sola_2017-2018} have been examined, such as a power series of $H$ \cite{Sola_2015L14}.
In particular, the simple combination of the constant and $H^{2}$ terms, i.e., 
$C_{0} H_{0}^{2} + C_{1} H^{2}$,
has been extensively examined and has been found to be favored, where $C_{0}$ and $C_{1}$ are dimensionless constants.
See, e.g., the works of Sol\`{a} \textit{et al.} \cite{Sola_2015L14},  Lima \textit{et al.} \cite{LimaSola_2013a}, and G\'{o}mez-Valent \textit{et al.} \cite{Valent2015}.
In addition, observations have revealed that $C_{1}$ is small, whereas $C_{0}$ for the constant term is dominant \cite{Sola_2015L14}.
Therefore, it is expected that we should focus on the constant term in discussing the order of the driving terms. 
The constant term is written as
\begin{equation}
  f_{cst} (t) = C_{0} H_{0}^{2}   .
\label{LCDM_cst}
\end{equation} 
In the $\Lambda (t)$CDM model, the constant term can be obtained from an integral constant of the renormalization group equation for the vacuum energy density \cite{Sola_2013}.

In the present study, a holographic equipartition model is assumed to be a particular case of $\Lambda (t)$CDM models, although the theoretical backgrounds are different \cite{Koma9,Koma10,Koma11}.
For example, from Eqs.\ (\ref{eq:L(t)_FRW01}) and (\ref{eq:L(t)_FRW02}), the continuity equation for the $\Lambda (t)$CDM model \cite{Koma9,Koma10} is given by
\begin{equation}
       \dot{\rho}(t) + 3  \frac{\dot{a}(t)}{a(t)} \left (  \rho (t) + \frac{p(t)}{c^2}  \right )    =  - \frac{ 3 \dot{f}(t)}{8 \pi G}        .
\label{eq:drho_L(t)}
\end{equation}
The right-hand side of this equation is non-zero, except for the case of a constant driving term.
This nonzero right-hand side implies a kind of energy exchange cosmology in which the transfer of energy between two fluids is assumed \cite{Barrow22}. 
Based on the holographic principle, the nonzero right-hand side can be interpreted as a kind of transfer of energy between the bulk (the universe) and the boundary (the horizon of the universe) \cite{Koma10}.

\section{Bekenstein--Hawking entropy  $S_{\rm{BH}}$ on the horizon of the universe} 
\label{Bekenstein-Hawking entropy}

The Bekenstein--Hawking entropy \cite{Bekenstein1Hawking1} is considered to be the standard scale of an associated entropy on the horizon of the universe.
In this section, the Bekenstein--Hawking entropy is reviewed, according to Ref.\ \cite{Koma11}.
We consider an entropy on the Hubble horizon of a flat FRW universe, in which an apparent horizon is equivalent to the Hubble horizon \cite{Easson12}.

The Bekenstein--Hawking entropy $S_{\rm{BH}}$ is written as
\begin{equation}
 S_{\rm{BH}}  = \frac{ k_{B} c^3 }{  \hbar G }  \frac{A_{H}}{4}   ,
\label{eq:SBH}
\end{equation}
where $k_{B}$ and $\hbar$ are the Boltzmann constant and the reduced Planck constant, respectively \cite{Bekenstein1Hawking1}. 
The reduced Planck constant is defined as $\hbar \equiv h/(2 \pi)$, where $h$ is the Planck constant \cite{Koma10,Koma11}.
Here, $A_{H}$ is the surface area of the sphere with the Hubble horizon (radius) $r_{H}$ given by
\begin{equation}
     r_{H} = \frac{c}{H}   .
\label{eq:rH}
\end{equation}
Substituting $A_{H}=4 \pi r_{H}^2 $ into Eq.\ (\ref{eq:SBH}) and using Eq.\ (\ref{eq:rH}), we obtain \cite{Koma9,Koma10,Koma11} 
\begin{equation}
S_{\rm{BH}}  = \frac{ k_{B} c^3 }{  \hbar G }   \frac{A_{H}}{4}       
                  =  \left ( \frac{ \pi k_{B} c^5 }{ \hbar G } \right )  \frac{1}{H^2}  
                  =    \frac{K}{H^2}    , 
\label{eq:SBH2}      
\end{equation}
where $K$ is a positive constant given by
\begin{equation}
  K =  \frac{  \pi  k_{B}  c^5 }{ \hbar G } = \frac{  \pi  k_{B}  c^2 }{ L_{p}^{2} }  , 
\label{eq:K-def}
\end{equation}
and $L_{p}$ is the Planck length, which is written as
\begin{equation}
  L_{p} = \sqrt{ \frac{\hbar G} { c^{3} } } .
\label{eq:Lp}
\end{equation}
Using Eq.\ (\ref{eq:SBH2}) and $H \neq 0$ \cite{PERL1998_Riess1998,Planck2015,Riess2016}, we have a positive entropy:
\begin{equation}
S_{\rm{BH}}  =    \frac{K}{H^2}    > 0 .
\label{eq:SBH2_plus}      
\end{equation}
Differentiating Eq.\ (\ref{eq:SBH2}) with respect to $t$, we obtain the rate of change of entropy, written as \cite{Koma11}
\begin{equation}
\dot{S}_{\rm{BH}}  =  \frac{d}{dt} \left ( \frac{K}{H^{2}} \right )  =  \frac{-2K \dot{H} }{H^{3}}   .  
\label{eq:dSBH}      
\end{equation}
Various observations indicate that $H >0$ and $\dot{H} < 0$ \cite{Krishna20172018}, see, e.g., Ref.\ \cite{Farooq2017}.
Accordingly, the second law of thermodynamics for the Bekenstein--Hawking entropy should satisfy \cite{Koma11}
\begin{equation}
\dot{S}_{\rm{BH}}   =  \frac{-2K \dot{H} }{H^{3}}  >  0  .  
\label{eq:dSBH_2}      
\end{equation}
In our universe, we assume that  $\dot{S}_{\rm{BH}}  > 0$.

\section{Holographic equipartition model with an arbitrary entropy $S_{H}$} 
\label{Holographic equipartition}

In this section, a holographic equipartition model is introduced, in accordance with previous studies \cite{Koma10,Koma11}, based on the original work of Padmanabhan \cite{Padma2012A}, and other related research \cite{Padma2012,Padma2014-2015,Cai2012-Tu2013,Yuan2013,Tu2013-2015,Krishna20172018,Neto2018a,Hadi-Sheykhia2018}.   
Although the assumption of equipartition of energy used for this model has not yet been established in a cosmological spacetime \cite{Koma11}, we herein assume the scenario to be viable.
In addition, we assume an arbitrary entropy $S_{H}$ on the horizon, as an alternative to the Bekenstein--Hawking entropy.
The arbitrary entropy is discussed later.

In an infinitesimal interval $dt$ of cosmic time, the increase $dV$ of the cosmic volume can be expressed as 
\begin{equation}
     \frac{dV}{dt}  =  L_{p}^{2} (N_{\rm{sur}} - \epsilon N_{\rm{bulk}} ) \times c      , 
\label{dVdt_N-N}
\end{equation}
where $N_{\rm{sur}}$ is the number of degrees of freedom on a spherical surface of Hubble radius $r_{H}$, whereas $N_{\rm{bulk}}$ is the number of degrees of freedom in the bulk \cite{Padma2012A}. 
The term $L_{p}$ is the Planck length given by Eq.\ (\ref{eq:Lp}), and $\epsilon$ is a parameter discussed below.
The right-hand side of Eq.\ (\ref{dVdt_N-N}) includes $c$, because $c$ is not set to be $1$ herein \cite{Koma10,Koma11}. 

Equation\ (\ref{dVdt_N-N}) comes from the work of Padmanabhan \cite{Padma2012A}.
The left-hand side of Eq.\ (\ref{dVdt_N-N}) represents the expansion of cosmic space, whereas the right-hand side represents the difference between the degrees of freedom on the surface and in the bulk. 
Accordingly, Eq.\ (\ref{dVdt_N-N}) indicates that the difference between the degrees of freedom is assumed to lead to the expansion of cosmic space \cite{Padma2012A}. 
This is the so-called holographic equipartition law proposed by Padmanabhan and has been examined from various viewpoints \cite{Padma2012,Padma2014-2015,Cai2012-Tu2013,Yuan2013,Tu2013-2015,Krishna20172018,Neto2018a,Hadi-Sheykhia2018,Koma10,Koma11}.  
In the present paper, we assume Eq.\ (\ref{dVdt_N-N}) to be viable although it has not yet been established.

The acceleration equation can be derived from Eq.\ (\ref{dVdt_N-N}), as examined in Ref.\ \cite{Padma2012A}.
In order to derive the acceleration equation, we first calculate the left-hand side of Eq.\ (\ref{dVdt_N-N}).
Using $V= \frac{4 \pi}{3} r_{H}^{3}$ and $r_{H}= c/H$, the rate of change of the Hubble volume $V$ is given by \cite{Padma2012A,Koma10,Koma11}
\begin{equation}
     \frac{dV}{dt}  =    \frac{d}{dt} \left [ \frac{4 \pi}{3} \left ( \frac{c}{H} \right )^{3}   \right ]  
                         =  -4 \pi c^{3}   \left (  \frac{ \dot{H} }{H^{4} } \right )  .
\label{dVdt_left}
\end{equation}
Next, we calculate the right-hand side of Eq.\ (\ref{dVdt_N-N}). 
To this end, parameters included in the right-hand side are introduced.
The number of degrees of freedom in the bulk is assumed to obey the equipartition law of energy \cite{Padma2012A}: 
\begin{equation}
  N_{\rm{bulk}} = \frac{|E|}{ \frac{1}{2} k_{B} T}     , 
\label{N_bulk}
\end{equation}
where the Komar energy $|E|$ contained inside the Hubble volume $V$ is given by 
\begin{equation}
|E| =  |( \rho c^2 + 3p)| V  = - \epsilon ( \rho c^2 + 3p) V  ,
\label{Komar}
\end{equation}
and $\epsilon$ is a parameter defined as \cite{Padma2012A,Padma2012}
 \begin{equation}
        \epsilon \equiv     
 \begin{cases}
              +1  & (\rho c^2 + 3p <0  \textrm{: an accelerating universe}),  \\ 
              -1  & (\rho c^2 + 3p >0   \textrm{: a decelerating universe}).    \\
\end{cases}
\label{epsilon}
\end{equation}
In the present paper, $\rho c^2 + 3p <0$ is selected, and, therefore, $\epsilon = +1$ from Eq.\ (\ref{epsilon}).
The acceleration equation discussed below is not affected by this selection \cite{Padma2012,Koma10,Koma11}.
The temperature $T$ on the horizon is written as
\begin{equation}
 T = \frac{ \hbar H}{   2 \pi  k_{B}  }   .
\label{eq:T0}
\end{equation}
The number of degrees of freedom on the spherical surface is given by  
\begin{equation}
  N_{\rm{sur}} = \frac{4 S_{H} }{k_{B}}       , 
\label{N_sur}
\end{equation}
where $S_{H}$ is the entropy on the Hubble horizon \cite{Koma10,Koma11}. 
When $S_{H} = S_{\rm{BH}}$, Eq.\ (\ref{N_sur}) is equivalent to that in Ref.\ \cite{Padma2012A}. 
In the present study, an arbitrary entropy $S_{H}$ on the horizon is assumed in order to discuss various types of entropy.
Keep in mind that $S_{H} > 0$ is considered herein. 
Further constraints on $S_{H}$ are discussed in Section\ \ref{SL}.

We now calculate the right-hand side of Eq.\ (\ref{dVdt_N-N}). 
Using $\epsilon = +1$, Eqs.\ (\ref{eq:Lp}), (\ref{N_bulk}), (\ref{Komar}), (\ref{eq:T0}), and (\ref{N_sur}), and calculating several operations \cite{Koma10}, the right-hand side of Eq.\ (\ref{dVdt_N-N}) can be written as 
\begin{align}
  &  L_{p}^{2} (N_{\rm{sur}} - \epsilon N_{\rm{bulk}} )  \times c      \notag \\
                                                                                         &=          \frac{\hbar G} { c^{3} }  \left [ \frac{4 S_{H} }{k_{B}} + \frac{ (4 \pi)^{2} c^{5}  }{ 3 \hbar } \left (  \rho  + \frac{3p}{c^{2}}  \right )  \frac{1}{  H^{4}   }  \right ] \times c     .
\label{N-N_right}
\end{align}
Equations\ (\ref{dVdt_left}) and (\ref{N-N_right}) are the left-hand and right-hand sides of Eq.\ (\ref{dVdt_N-N}), respectively.
Accordingly, from Eqs.\ (\ref{dVdt_left}) and (\ref{N-N_right}), Eq.\ (\ref{dVdt_N-N}) is written as \cite{Koma10}
\begin{equation}
 \frac{ -4 \pi c^{3}   \dot{H} }{H^{4} }   = \frac{\hbar G} { c^{3} }  \left [ \frac{4 S_{H} }{k_{B}} + \frac{ (4 \pi)^{2} c^{5}  }{ 3 \hbar } \left (  \rho  + \frac{3p}{c^{2}}  \right )  \frac{1}{  H^{4}   }  \right ] \times c     .
\label{dVdt_N-N_cal1}
\end{equation}
Solving this equation with regard to $\dot{H}$, we have 
\begin{align}
  \dot{H}  &=  - \frac{ H^{4} }{ 4 \pi c^{3} }   \frac{\hbar G} { c^{3} }  \left [ \frac{4 S_{H} }{k_{B}} +  \frac{ (4 \pi )^{2} c^{5}  }{ 3 \hbar } \left (  \rho  + \frac{3p}{c^{2}}  \right )  \frac{1}{  H^{4}   }  \right ]  \times c   \notag \\
               &=   -  \frac{ 4 \pi G }{ 3} \left (  \rho  + \frac{3p}{c^{2}}  \right )  - \left ( \frac{ \hbar G }{  \pi k_{B} c^{5} } \right )  S_{H} H^{4}                                                                                                             \notag \\
               &=   -  \frac{ 4 \pi G }{ 3} \left (  \rho  + \frac{3p}{c^{2}}  \right )  - \frac{ S_{H} H^{4} }{K}                                                                                                    , 
\label{dVdt_N-N_cal2}
\end{align}
where $K$ is given by Eq.\ (\ref{eq:K-def}).
As shown in Eq.\ (\ref{eq:L(t)_FRW02}), $\ddot{a}/ a$ is written as $ \ddot{a}/ a   =  \dot{H} + H^{2}$.
Using this and Eq.\ (\ref{dVdt_N-N_cal2}), we have the following acceleration equation \cite{Koma10}: 
\begin{equation}
  \frac{ \ddot{a} }{ a }       =  \dot{H} + H^{2}     =   -  \frac{ 4 \pi G }{ 3} \left (  \rho  + \frac{3p}{c^{2}}  \right )  - \frac{ S_{H} H^{4} }{K}      + H^{2}    .
\label{N-N_FRW02_SH_000}
\end{equation}
This equation can be arranged using the Bekenstein--Hawking entropy, which is considered to be the standard scale of the entropy on the horizon.
From Eq.\ (\ref{N-N_FRW02_SH_000}) and $S_{\rm{BH}} = K/H^{2}$ given by Eq.\ (\ref{eq:SBH2}),  
the acceleration equation for the holographic equipartition model is \cite{Koma11}
\begin{align}
  \frac{ \ddot{a} }{ a }       
                                      &=   -  \frac{ 4 \pi G }{ 3} \left (  \rho  + \frac{3p}{c^{2}}  \right ) + H^{2}  \left (  \frac{S_{\rm{BH}} - S_{H} }{ S_{\rm{BH}} }  \right )    \notag \\ 
                                      &= -  \frac{ 4 \pi G }{ 3} \left (  \rho  + \frac{3p}{c^{2}}  \right ) + f_{h} (t)  ,
\label{N-N_FRW02_SH_2}
\end{align}
where an extra driving term $f_{h} (t)$ is given by
\begin{equation}
f_{h} (t)  = H^{2} \left (  \frac{S_{\rm{BH}} - S_{H} }{ S_{\rm{BH}} }  \right ) = H^{2}  \left (  1 - \frac{ S_{H} }{ S_{\rm{BH}} }  \right )  .
\label{eq:f1(t)}
\end{equation}
This equation indicates that a deviation of $S_{H}$ from $S_{\rm{BH}}$ plays an important role in the driving term.
(Note that the entropy on the horizon is assumed to be $S_{H}$ larger than zero.)
For example, when $S_{H} = S_{\rm{BH}}$, the driving term $f_{h} (t)$ is zero \cite{Padma2012A}.
However, $f_{h} (t)$ is non-zero when $S_{H} \neq S_{\rm{BH}}$ \cite{Koma10,Koma11}.
In particular, when $S_{H} < S_{\rm{BH}}$, a positive driving term for an accelerating universe is obtained from Eq.\ (\ref{eq:f1(t)}).
That is, the deviation from $S_{\rm{BH}}$ plays important roles in the holographic equipartition model. 
This result may imply that such an effective entropy is considered to be favored, as an alternative to $S_{\rm{BH}}$.
In fact, a modified R\'{e}nyi entropy \cite{Czinner1} and a power-law corrected entropy \cite{Das2008} have been used for $S_{H}$ \cite{Koma10,Koma11}.
We briefly review typical examples in the next paragraph.

By regarding the Bekenstein--Hawking entropy as a nonextensive Tsallis entropy \cite{Tsa0} and using a logarithmic formula \cite{Koma10}, a modified R\'{e}nyi entropy suggested by Bir\'{o} and Czinner \cite{Czinner1} is a novel type of R\'{e}nyi entropy \cite{Ren1} on horizons. 
The modified R\'{e}nyi entropy $S_{R}$ is given by 
\begin{equation}
 S_{R} = \frac{1}{\lambda} \ln [1+ \lambda S_{\rm{BH}} ] , 
\end{equation}
where $\lambda=1-q$, and $q$ is a non-extensive parameter.
By substituting $S_{H}=S_{R}$ into Eq.\ (\ref{eq:f1(t)}) and calculating several operations \cite{Koma10}, the extra driving term $f_{h,R}(t)$ can be written as 
\begin{equation}
 f_{h,R}(t)  = H^{2}  \left (  1 - \frac{  \ln [1+ (\lambda K /H^{2}) ]  }{  \lambda K /H^{2} }  \right )      .
\end{equation}
In contrast, a power-law corrected entropy suggested by Das \textit{et al.} \cite{Das2008} is based on the entanglement of quantum fields between the inside and the outside of the horizon \cite{Koma11}.
The power-law corrected entropy $S_{pl}$ \cite{Radicella2010} can be written as  
\begin{equation}
 S_{pl}  = S_{\rm{BH}}  \left [ 1-  \Psi_{\alpha} \left ( \frac{H_{0}}{H} \right )^{2- \alpha}  \right ] , 
\end{equation}
where $\alpha$ and $\Psi_{\alpha}$ are dimensionless positive parameters related to the entanglement \cite{Koma11}.
Substituting $S_{H}=S_{pl}$ into Eq.\ (\ref{eq:f1(t)}) and calculating several operations \cite{Koma11}, the driving term $f_{h,pl}(t)$ can be written as 
\begin{equation}
        f_{h,pl} (t) =  \Psi_{\alpha} H_{0}^{2} \left (  \frac{H}{H_{0}} \right )^{\alpha}  . 
\end{equation}
The two driving terms, i.e.,  $f_{h,R}(t)$ and $f_{h,pl} (t)$, tend to be constant-like when $\lambda K/H^{2} < 1$ and $\alpha <1$.
For details, see Refs.\ \cite{Koma10,Koma11}.

As shown in Eq.\ (\ref{eq:f1(t)}), the deviation of $S_{H}$ from $S_{\rm{BH}}$ can lead to an extra driving term $f_{h} (t)$.
In the holographic equipartition model, $f_{h} (t)$ is expected to be constrained by the second law of thermodynamics because $f_{h} (t)$ depends on $S_{H}$.
In Section\ \ref{SL}, we will discuss the order of $f_{h} (t)$ from a thermodynamics viewpoint.

Apart from the holographic equipartition model, Padmanabhan reported that the observed cosmological constant can arise from vacuum fluctuations (or modifications) of energy density rather than the vacuum energy itself  \cite{Padma2005}.
The vacuum fluctuations may be related to the deviation of entropy considered herein.

Recently, $\Lambda (t)$CDM models have been closely examined \cite{Sola_2017-2018}, in order to resolve current problems, such as a significant tension between the Planck 2015 results \cite{Planck2015} and the local (distance ladder) measurement from the Hubble Space Telescope \cite{Riess2016}.
Consequently, a certain type of $\Lambda (t)$CDM models has been found to be favored as compared to $\Lambda$CDM models \cite{Sola_2017-2018}.
As described in Section\ \ref{LCDM models}, cosmological equations in the holographic equipartition model are expected to be similar to those for $\Lambda (t)$CDM models.
Therefore, if a particular entropy on the horizon can be assumed, the obtained holographic equipartition model should be favored, as for the $\Lambda (t)$CDM model.
Detailed studies are needed and, this task is left for future research.

\section{Second law of thermodynamics for the present model} 
\label{SL}

In the previous section, we derived an extra driving term in a holographic equipartition model.
In this section, we examine thermodynamic constraints on the driving term using both the positivity of entropy and the second law of thermodynamics.
In the present study, an arbitrary entropy $S_{H}$ on the horizon is assumed, extending previous works \cite{Koma10,Koma11}.
Accordingly, we can universally examine the thermodynamic constraints.

To discuss the generalized second law of thermodynamics, we consider the total entropy $S_{t}$ given by
\begin{equation}
      S_{t} = S_{H} + S_{m}    \quad   \textrm{and, therefore,} \quad    \dot{S}_{t} = \dot{S}_{H} + \dot{S}_{m} , 
\label{eq:S_t0_12}
\end{equation}
where $S_{m}$ is the entropy of matter inside the horizon \cite{Koma11}.
In the present study, the holographic equipartition model is assumed to be a particular case of $\Lambda (t)$CDM models.
Therefore, $\dot{S}_{m}$ for the model is the same as $\dot{S}_{m}$ examined in a previous study \cite{Koma11}.

According to Ref.\ \cite{Koma11}, $ \dot{S}_{m}$ can be calculated from the first law of thermodynamics.
From Eq.\ (A5) in the Appendix of Ref.\ \cite{Koma11}, we have $\dot{S}_{m}$, which is written as
\begin{equation}
\dot{S}_{m}  =  \frac{ - \dot{f}_{h} (t) K}{ H^4 }  , 
\label{eq:dSm_a_12}      
\end{equation}
where $f_{h} (t)$ is an extra driving term for the holographic equipartition model.
For details, see Ref.\ \cite{Koma11}.
This equation can be arranged using the Bekenstein--Hawking entropy.
Using $\dot{S}_{\rm{BH}}$ given by Eq.\ (\ref{eq:dSBH}), $\dot{S}_{m}$ is written as 
\begin{equation}
\dot{S}_{m}  =  \frac{ - \dot{f}_{h} (t) K}{ H^4 } = \frac{-2K \dot{H} }{H^{3}} \frac{  \dot{f}_{h} (t)}{ 2 H \dot{H} }   =  \frac{ \dot{S}_{\rm{BH}}  \dot{f}_{h} (t)}{ 2 H \dot{H} }   .
\label{eq:dSm_b_12}      
\end{equation}

We now examine thermodynamic constraints on an extra driving term $f_{h} (t)$ in the holographic equipartition model.
From Eq.\ (\ref{eq:f1(t)}), the driving term is written as
\begin{equation}
f_{h} (t)  = H^{2} \left (  \frac{S_{\rm{BH}} - S_{H} }{ S_{\rm{BH}} }  \right ) = H^{2}  \left (  1 - \frac{ S_{H} }{ S_{\rm{BH}} }  \right )  .
\label{eq:f1(t)_2}
\end{equation}
Solving this equation with regard to $S_{H}$ gives
\begin{equation}
 S_{H}  =  S_{\rm{BH}}  \left ( 1 - \frac{ f_{h} (t) }{H^2} \right )    .
\label{eq:SH1}
\end{equation}

Before discussing the second law of thermodynamics, we examine the positivity of $S_{H}$.
From Eq.\ (\ref{eq:SBH2_plus}), we have $S_{\rm{BH}} >  0$.
Accordingly, in order to satisfy $S_{H} >  0 $, we require
\begin{equation}
  1 - \frac{ f_{h} (t) }{H^2}  > 0       ,
\label{eq:SH1_positive_1}
\end{equation}
or equivalently,
\begin{equation}
    f_{h} (t)   < H^2    .
\label{eq:SH1_positive_2}
\end{equation}
The inequalities given by Eqs.\ (\ref{eq:SH1_positive_1}) and (\ref{eq:SH1_positive_2}) imply an upper limit of $f_{h} (t)$.
Numerous observations indicate $\dot{H}<0$ \cite{Krishna20172018} and, therefore, $H^2$ is a minimum when $H= H_{0}$.
Thus, the strictest constraint can be written as
\begin{equation}
 f_{h} (t)     <   H_{0}^{2}    \quad  ( \leq H^{2} )    .
\label{SH1_positive_H0_1}
\end{equation}
From Eq.\ (\ref{SH1_positive_H0_1}), the order of the extra driving term, $f_{h} (t)$, can be approximately written as
\begin{equation}
  O  ( f_{h} (t)  )  \lessapprox     O \left (  H_{0}^{2}   \right )       .
\label{SH1_positive_order}
\end{equation}
The positivity of $S_{H}$ is found to constrain $f_{h} (t)$.
In addition, the order of $f_{h} (t)$ is consistent with the order of the observed $\Lambda$ given by Eq.\ (\ref{L_order_0}).

Now, we discuss the generalized second law of thermodynamics.
To this end, we first calculate the rate of change of $S_{H}$.
Differentiating Eq.\ (\ref{eq:SH1}) with respect to $t$ and using Eqs.\ (\ref{eq:SBH2}) and (\ref{eq:dSBH}), $\dot{S}_{H}$ can be written as
\begin{align}
\dot{S}_{H}  
                 &=  \dot{S}_{\rm{BH}}  \left ( 1 - \frac{ f_{h} (t) }{H^2} \right )    + {S}_{\rm{BH}}  \frac{d}{dt} \left ( 1 - \frac{ f_{h} (t) }{H^2} \right )   \notag \\
                 &=  \frac{-2K \dot{H} }{H^{3}} \left ( 1 - \frac{    f_{h} (t) }{H^2} \right )  +  \frac{K}{H^{2}}  \left (  - \frac{ \dot{f}_{h} (t) }{H^2} +  \frac{ 2 f_{h} (t)  \dot{H} }{H^3} \right )   \notag \\
                 &=  \frac{-2K \dot{H} }{H^{3}} \left ( 1 - \frac{ 2 f_{h} (t) }{H^2} +  \frac{ \dot{f}_{h} (t) }{ 2H \dot{H} } \right )   \notag \\
                 &=  \dot{S}_{\rm{BH}}  \left ( 1 - \frac{ 2 f_{h} (t) }{H^2} +  \frac{ \dot{f}_{h} (t) }{ 2H \dot{H} } \right )        .
\label{eq:dSH1}      
\end{align}
Using Eqs.\ (\ref{eq:dSH1}) and (\ref{eq:dSm_b_12}), the rate of change of the total entropy in the holographic equipartition model is given by
\begin{equation}
      \dot{S}_{t,h}  = \dot{S}_{H} + \dot{S}_{m}    = \dot{S}_{\rm{BH}}  \left ( 1 - \frac{ 2 f_{h} (t) }{H^2} +  \frac{ \dot{f}_{h} (t) }{ H \dot{H} } \right )    ,
\label{eq:dS_t1}
\end{equation} 
where $\dot{S}_{\rm{BH}} >  0$ from Eq. (\ref{eq:dSBH_2}).
In order to satisfy $\dot{S}_{t,h} >  0 $, we require
\begin{equation}
 1 - \frac{ 2 f_{h} (t) }{H^2} +  \frac{ \dot{f}_{h} (t) }{ H \dot{H} }   >0      ,
\label{dSt_ineq_1}
\end{equation} 
or equivalently,
\begin{equation}
  f_{h} (t) <  \frac{H^2}{2} +  \frac{ \dot{f}_{h} (t) H }{ 2 \dot{H} }        .
\label{dSt_ineq_2}      
\end{equation}
The inequalities given by Eqs.\ (\ref{dSt_ineq_1}) and (\ref{dSt_ineq_2}) indicate that the extra driving term $f_{h} (t)$ is restricted by the generalized second law of thermodynamics.
These inequalities imply an upper limit of $f_{h} (t)$.
Note that a similar constraint can be obtained from $\dot{S}_{H} >0$, using Eq.\ (\ref{eq:dSH1}).

Let us examine a typical constant driving term corresponding to $\Lambda$CDM models, because the constant term is dominant in $\Lambda(t)$CDM models \cite{Sola_2015L14}, as described in Section\ \ref{LCDM models}.
To this end, we consider $f_{h}(t) \approx f_{h,cst} (t) = C_{0} H_{0}^{2}$ given by Eq.\ (\ref{LCDM_cst}), where $C_{0}$ is a dimensionless constant.
In the holographic equipartition model, the constant term can be obtained from an entropy given by $S_{H} = S_{\rm{BH}} [1 - C_{0} H_{0}^{2} H^{-2}] $.
This entropy is equivalent to a power-law corrected entropy for a small entanglement, corresponding to $\alpha \ll 1$ \cite{Koma11}.
Using Eq.\ (\ref{dSt_ineq_2}) and $\dot{f}_{h,cst} (t) = 0$, we have the strictest constraint, which is given by
\begin{equation}
 f_{h,cst} (t)     <   \frac{H_{0}^{2}}{2}  \quad \left ( \leq \frac{H^{2}}{2} \right )  ,
\label{dSt_ineq_H0}
\end{equation}
where $H_{0} \leq H$ is also used. 
Accordingly, the order of the constant driving term can be approximately written as
\begin{equation}
  O  ( f_{h,cst} (t)  )  \lessapprox     O \left (  H_{0}^{2}   \right )       .
\label{dSt_ineq_H0_order}
\end{equation}
The constraint on $f_{h,cst} (t)$ agrees with Eq.\ (\ref{SH1_positive_order}), which is based on the positivity of $S_{H}$.
The order of $f_{h,cst} (t)$ is consistent with the order of the observed $\Lambda$ from Eq.\ (\ref{L_order_0}).
Therefore, the order of $f_{h} (t)$ is also consistent with the order of the observed $\Lambda$ because, as mentioned above, $f_{h}(t) \approx f_{h,cst} (t)$ is expected.

\begin{figure} [t]  
\begin{minipage}{0.49\textwidth}
\begin{center}
\scalebox{0.32}{\includegraphics{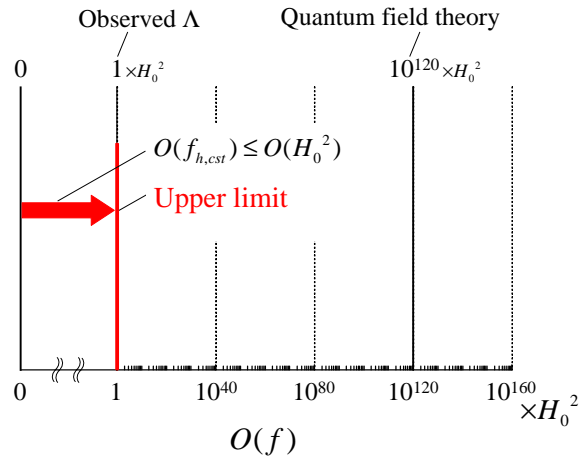}}
\end{center}
\end{minipage}
\caption{ (Color online). Thermodynamic constraints on the constant driving term for a holographic equipartition model.
The constraint on the constant term, i.e., $f_{h,cst}$, is given by Eq.\ (\ref{dSt_ineq_H0_order}).
The bold line with an arrow represents an allowed region corresponding to the thermodynamic constraint.
The orders of the observed $\Lambda$ and the theoretical value from quantum field theory are also shown.}
\label{Fig1r}
\end{figure}

The obtained constraint on $f_{h,cst} (t)$ is indicated in Fig.\ \ref{Fig1r}.
We can confirm an upper limit of $f_{h,cst} (t)$. 
The upper limit is consistent with the order of the observed $\Lambda$.
In this way, we can discuss thermodynamic constraints on extra driving terms, which are derived from various types of entropy on the horizon.
(In addition, we can examine constraints on $S_{H}$. 
For example, substituting Eq.\ (\ref{eq:f1(t)_2}) into Eq.\ (\ref{dSt_ineq_H0}) and using $S_{\rm{BH}} >0$ given by Eq.\ (\ref{eq:SBH2_plus}), we have $S_{H} > S_{\rm{BH}}/2 >0$.)

In the present study, several quantities have been systematically arranged using the Bekenstein--Hawking entropy.
In these quantities, the extra driving term and time derivatives of entropies are summarized in Table\ \ref{tab-summary_r}.
As shown in Table\ \ref{tab-summary_r}, the driving term $f_{h} (t)$ includes a relative difference of entropy, i.e., $(S_{\rm{BH}} -  S_{H}) / S_{\rm{BH}}$, due to the difference between the degrees of freedom on the surface and in the bulk.
The difference in entropy, i.e., $\Delta S = S_{\rm{BH}} -  S_{H}$, can lead to an extra driving term and its upper limit.
In cosmological scenarios based on the holographic principle, the difference in entropy may be interpreted as a kind of vacuum fluctuation (or modification) of energy density \cite{Padma2005}. 
Further studies are needed, and this task is left for future research.

\begin{table}[t]
\caption{Extra driving term and time derivatives of entropies derived from a holographic equipartition model. 
An associated entropy on the horizon is assumed to be $S_{H}$ larger than zero. 
The Bekenstein--Hawking entropy, $S_{\rm{BH}}$ given by Eq.\ (\ref{eq:SBH2}), is used for the standard scale of the entropy on the horizon.
}
\label{tab-summary_r}
\newcommand{\m}{\hphantom{$-$}}
\newcommand{\cc}[1]{\multicolumn{1}{c}{#1}}
\renewcommand{\tabcolsep}{1.5pc} 
\renewcommand{\arraystretch}{2.0} 
\begin{tabular}{@{}lllll}
\hline
\hline
$\textrm{Parameter}$     &    $\textrm{Holographic equipartition model}$                                                           \\
\hline
$f_{h}(t)$                       &   $  H^{2}  \left [  (S_{\rm{BH}} -  S_{H}) / S_{\rm{BH}}   \right ]   $                               \\
$\dot{S}_{H}$                 &   $ \dot{S}_{\rm{BH}}  \left [ 1 -  2 f_{h} (t) / H^2  +   \dot{f}_{h} (t) / (2H \dot{H} ) \right ] $
                                                                                                                                                                    \\
$\dot{S}_{t}$                  &   $ \dot{S}_{\rm{BH}}  \left [ 1 -  2 f_{h} (t) / H^2  +   \dot{f}_{h} (t) / ( H \dot{H} ) \right ]  $
                                                                                                                                                                     \\
\hline
\hline
\end{tabular}\\
 \end{table}

\section{Conclusions}
\label{Conclusions}

Holographic equipartition models are based on the holographic principle.
In the present study, we have assumed an arbitrary entropy $S_{H}$ on the horizon of the universe and applied this entropy to the holographic equipartition model in order to universally examine thermodynamic constraints on an extra driving term in a flat FRW universe at late times.
The driving term is systematically formulated using the Bekenstein--Hawking entropy, $S_{\rm{BH}}$, which is considered to be the standard scale of the entropy on the horizon of the universe.
Consequently, $H^{2}[( S_{\rm{BH}} - S_{H})/S_{\rm{BH}}]$ terms are obtained in the holographic equipartition model.
The formulation used here reveals that deviations of $S_{H}$ from $S_{\rm{BH}}$ play an essential role in the driving term.
In addition, the driving term is found to be constrained by the second law of thermodynamics, even if the arbitrary entropy is assumed.
(Note that $S_{H} > 0$ is considered herein.)
The second law of thermodynamics for the holographic equipartition model constrains the upper limit of the driving term.
The upper limit implies that the order of the driving term is likely consistent with the order of the cosmological constant measured by observations.
An approximately equivalent upper limit of the driving term can be derived from the positivity of $S_{H}$ in the holographic equipartition model. 

The present results may indicate that we can discuss the cosmological constant problem from a thermodynamics viewpoint, in the holographic equipartition model. 
Of course, all the other contributions such as quantum field theory cannot be excluded, because only the upper limit of the driving term is discussed in the present study. 
In addition, the assumption of equipartition of energy used in this model has not yet been established in a cosmological spacetime.
However, the thermodynamic constraints and systematic formulations examined herein should provide new insights into other cosmological models based on the holographic principle.

\begin{acknowledgements}
The present study is supported by JSPS KAKENHI Grant Number JP18K03613.
The author wishes to thank the anonymous referee for very valuable comments which improved the paper.
\end{acknowledgements}


\begin{thebibliography}{99}

\bibitem{PERL1998_Riess1998} S. Perlmutter \textit{et al.},  Nature (London) \textbf{391}, 51 (1998); A. G. Riess \textit{et al.}, Astron. J. \textbf{116}, 1009 (1998). 
\bibitem{Planck2015} P. A. R. Ade \textit{et al.}, Astron. Astrophys. \textbf{594}, A13 (2016).     
%
\bibitem{Riess2016}  A. G. Riess \textit{et al.}, Astrophys. J. \textbf{826}, 56 (2016).
%


\bibitem{Weinberg1989} 
S. Weinberg, Rev. Mod. Phys. \textbf{61}, 1 (1989);  
I. Zlatev, L. Wang, P. J. Steinhardt, Phys. Rev. Lett.  \textbf{82}, 896 (1999); 
V. Sahni, A. A. Starobinsky, Int. J. Mod. Phys. D \textbf{9}, 373 (2000); 
S. M. Carroll, Living Rev. Relativity \textbf{4}, 1 (2001); 
T. Padmanabhan, Phys. Rep. \textbf{380}, 235 (2003).



\bibitem{Bamba1} K. Bamba, S. Capozziello, S. Nojiri, S. D. Odintsov, Astrophys. Space Sci. \textbf{342}, 155 (2012).




\bibitem{Freese-Mimoso_2015}
K. Freese, F. C. Adams, J. A. Frieman, E. Mottola, Nucl. Phys. \textbf{B287}, 797 (1987); 
J. M. Overduin, F. I. Cooperstock, Phys. Rev. D \textbf{58}, 043506 (1998); 
I. L. Shapiro, J. Sol\`{a}, J. High Energy Phys. 02 (2002) 006; 
%
J. P. Mimoso, D. Pav\'{o}n, Phys. Rev. D \textbf{87}, 047302 (2013);
M. H. P. M. Putten, Mon. Not. R. Astron. Soc. \textbf{450}, L48 (2015).
%
\bibitem{Sola_2009-2015} 
S. Basilakos, M. Plionis, J. Sol\`{a}, Phys. Rev. D \textbf{80}, 083511 (2009);
J. Grande, J. Sol\`{a}, S. Basilakos, M. Plionis, J. Cosmol. Astropart. Phys. 08 (2011) 007; 
E. L. D. Perico, J. A. S. Lima, S. Basilakos, J. Sol\`{a}, Phys. Rev. D \textbf{88}, 063531 (2013); 
%
%
J. Sol\`{a}, Int. J. Mod. Phys. A \textbf{31}, 1630035 (2016); 
%
S. Basilakos, A. Paliathanasis, J. D. Barrow, G. Papagiannopoulos, Eur. Phys. J. C \textbf{78}, 684 (2018).
%
%
%
%
\bibitem{Nojiri2006}  
S. Nojiri, S. D. Odintsov, Phys. Lett. B \textbf{639}, 144 (2006); 
%
%
Y. Wang, D. Wands, G.-B. Zhao, L. Xu, Phys. Rev. D \textbf{90}, 023502 (2014); 
N. Tamanini, Phys. Rev. D \textbf{92}, 043524 (2015); 
Q. Wang, Z. Zhu, W. G. Unruh, Phys. Rev. D \textbf{95}, 103504 (2017).
%
\bibitem{Sola_2015L14}
J. Sol\`{a}, A. G\'{o}mez-Valent, J. C. P\'{e}rez, Astrophys. J. \textbf{811}, L14 (2015).
%
\bibitem{LimaSola_2013a} 
J. A. S. Lima, S. Basilakos, J. Sol\`{a},  Mon. Not. R. Astron. Soc. \textbf{431}, 923 (2013).
%
\bibitem{LimaSola_2015_2016} 
J. A. S. Lima, S. Basilakos, J. Sol\`{a}, Gen. Relativ. Gravit. \textbf{47}, 40 (2015);  
Eur. Phys. J. C \textbf{76}, 228 (2016).
%
%
\bibitem{Valent2015} 
A. G\'{o}mez-Valent, J. Sol\`{a}, S. Basilakos, J. Cosmol. Astropart. Phys. 01 (2015) 004.
%
\bibitem{Sola_2013} J. Sol\`{a}, J. Phys. Conf. Ser. \textbf{453},  012015 (2013).
%
%
%
\bibitem{Sola_2017-2018}
J. Sol\`{a}, A. G\'{o}mez-Valent, J. C. P\'{e}rez, Phys. Lett. B \textbf{774}, 317 (2017);
A. G\'{o}mez-Valent, J. Sol\`{a} Peracaula, Mon. Not. R. Astron. Soc. \textbf{478}, 126 (2018);
J. Sol\`{a} Peracaula, J. C. P\'{e}rez, A. G\'{o}mez-Valent, Mon. Not. R. Astron. Soc. \textbf{478}, 4357 (2018);
J. Sol\`{a} Peracaula, A. G\'{o}mez-Valent, J. C. P\'{e}rez, arXiv:1811.03505.






%
%
\bibitem{Prigogine_1988-1989}  
I. Prigogine, J. Geheniau, E. Gunzig, P. Nardone, Proc. Natl. Acad. Sci. U.S.A. \textbf{85}, 7428 (1988); 
Gen. Relativ. Gravit. \textbf{21}, 767 (1989).
%
\bibitem{Lima1992-1996}   
M. O. Calv\~{a}o, J. A. S. Lima, I. Waga, Phys. Lett. A \textbf{162}, 223 (1992);
J. A. S. Lima, A. S. M. Germano, L. R. W. Abramo, Phys. Rev. D \textbf{53}, 4287 (1996).
%
\bibitem{LimaOthers2001-2016}   
W. Zimdahl, D. J. Schwarz, A. B. Balakin, D. Pav\'{o}n, Phys. Rev. D \textbf{64}, 063501 (2001); 
J. A. S. Lima, S. Basilakos, F. E. M. Costa, Phys. Rev. D \textbf{86}, 103534 (2012); 
%
T. Harko, Phys. Rev. D \textbf{90}, 044067 (2014); 
J. A. S. Lima, R. C. Santos, J. V. Cunha, J. Cosmol. Astropart. Phys. 03 (2016) 027;
%
R. C. Nunes, Gen. Relativ. Gravit. \textbf{48}, 107 (2016).










\bibitem{Sheykhi1}
A. Sheykhi, Phys. Rev. D \textbf{81}, 104011 (2010);
K. Karami, A. Sheykhi, N. Sahraei, S. Ghaffari, Eur. Phys. Lett. \textbf{93}, 29002 (2011).
%
\bibitem{Sadjadi1} H. M. Sadjadi, M. Jamil, Eur. Phys. Lett. \textbf{92}, 69001 (2010); 
%
S. Mitra, S. Saha, S. Chakraborty, Mod. Phys. Lett. A \textbf{30}, 1550058 (2015).








\bibitem{Padma2012A}  T. Padmanabhan, arXiv:1206.4916 [hep-th].
\bibitem{Padma2012}  T. Padmanabhan, Res. Astron. Astrophys. \textbf{12}, 891 (2012).
%
\bibitem{Padma2014-2015}  
T. Padmanabhan, H. Padmanabhan, Int. J. Mod. Phys. D \textbf{23}, 1430011 (2014); 
T. Padmanabhan, Mod. Phys. Lett. A \textbf{30}, 1540007 (2015).


\bibitem{Cai2012-Tu2013}                                                                               
R. G. Cai, J. High Energy Phys. 1211 (2012) 016; 
K. Yang, Y.-X. Liu, Y.-Q. Wang, Phys. Rev. D \textbf{86} 104013 (2012); 
A. Sheykhi,  M. H. Dehghani, S. E. Hosseini, Phys. Lett. B \textbf{726}, 23 (2013).

\bibitem{Yuan2013} Fang-Fang Yuan, Yong-Chang Huang, arXiv:1304.7949v3 [gr-qc].

\bibitem{Tu2013-2015}  
Fei-Quan Tu, Yi-Xin Chen, J. Cosmol. Astropart. Phys. 05 (2013) 024; 
%
A. F. Ali, Phys. Lett. B \textbf{732}, 335 (2014);
%
WY Ai, H. Chen, XR. Hu, JB. Deng, Gen. Relativ. Gravit. \textbf{46}, 1680 (2014); 
%
%
Fei-Quan Tu, Yi-Xin Chen, Gen. Relativ. Gravit. \textbf{47}, 87 (2015);
%
%
S. Chakraborty, T. Padmanabhan, Phys. Rev. D \textbf{90}, 124017 (2014); 
%
S. Chakraborty, T. Padmanabhan, Phys. Rev. D \textbf{92}, 104011 (2015); 
%
H. Moradpour, Int. J. Theor. Phys. \textbf{55}, 4176 (2016);  
%
%
T. Bandyopadhyay, Advances in High Energy Physics \textbf{2018}, 3752641 (2018).
%


\bibitem{Krishna20172018} P. B. Krishna, T. K. Mathew, Phys. Rev. D \textbf{96}, 063513 (2017);  
arXiv:1805.01705v2 [gr-qc].



\bibitem{Hadi-Sheykhia2018} H. Hadi, Y. Heydarzade, M. Hashemi, F. Darabi, Eur. Phys. J. C \textbf{78}, 38 (2018); 
A. Sheykhi, Phys. Lett. B \textbf{785}, 118 (2018). 
%
\bibitem{Neto2018a} E. M. C. Abreu, J. A. Neto, A. C. R. Mendes, A. Bonilla, Europhysics. Lett. \textbf{121}, 45002 (2018).
%



\bibitem{Koma10}  N. Komatsu, Eur. Phys. J. C \textbf{77}, 229 (2017).
\bibitem{Koma11}  N. Komatsu, Phys. Rev. D \textbf{96}, 103507 (2017).





\bibitem{Sheykhi2} A. Sheykhi, S. H. Hendi, Phys. Rev. D \textbf{84}, 044023 (2011). 
%



\bibitem{Karami2011-2018} 
K. Karami, A. Abdolmaleki, Z. Safari, S. Ghaffari, J. High Energy Phys. 08 (2011) 150; 
A. Sheykhi, M. Jamil, Gen. Relativ. Gravit. \textbf{43}, 2661 (2011); 
K. Karami, S. Asadzadeh, A. Abdolmaleki, Z. Safari, Phys. Rev. D \textbf{88}, 084034 (2013);
P. Saha, U. Debnath, Eur. Phys. J. C \textbf{76}, 491 (2016); 
%
%
%
H. Moradpour, A. Bonilla, E. M. C. Abreu, J. A. Neto, Phys. Rev. D \textbf{96}, 123504 (2017); 
%
A. Iqbal, A. Jawad, Adv. High Ener. Phys. \textbf{2018}, 6139430 (2018).








\bibitem{Easson12}  D. A. Easson, P. H. Frampton, G. F. Smoot, Phys. Lett. B \textbf{696}, 273 (2011); Int. J. Mod. Phys. A \textbf{27}, 1250066 (2012).
\bibitem{Koivisto1Basilakos1-Gohar}
Y. F. Cai, J. Liu, H. Li,      Phys. Lett. B \textbf{690}, 213 (2010); 
%
Y. F. Cai, E. N. Saridakis,  Phys. Lett. B \textbf{697}, 280 (2011); 
%
T. S. Koivisto, D. F. Mota, M. Zumalac\'{a}rregui, J. Cosmol. Astropart. Phys. 02 (2011) 027; 
T. Qiu, E. N. Saridakis, Phys. Rev. D \textbf{85}, 043504 (2012). 
S. Basilakos, D. Polarski, J. Sol\`{a}, Phys. Rev. D \textbf{86}, 043010 (2012); 
%
R. C. Nunes, E. M. Barboza Jr., E. M. C. Abreu, J. A. Neto, J. Cosmol. Astropart. Phys. 08 (2016) 051; 
%
%
I. D\'{i}az-Salda\~{n}a, J. L\'{o}pez-Dom\'{i}nguez, M. Sabido, arXiv:1806.04918.
%



\bibitem{Sola_2014a} 
S. Basilakos, J. Sol\`{a}, Phys. Rev. D \textbf{90}, 023008 (2014).
\bibitem{Gohar_2015a} M. P. D\c{a}browski, H. Gohar, Phys. Lett. B \textbf{748}, 428 (2015).
\bibitem{Gohar_2015b} 
M. P. D\c{a}browski, H. Gohar, V. Salzano, Entropy \textbf{18}, 60 (2016).


\bibitem{Koma4}  N. Komatsu, S. Kimura, Phys. Rev. D \textbf{87}, 043531 (2013); N. Komatsu, JPS Conf. Proc. \textbf{1}, 013112 (2014).
\bibitem{Koma5}  N. Komatsu, S. Kimura, Phys. Rev. D \textbf{88}, 083534 (2013).
\bibitem{Koma6}  N. Komatsu, S. Kimura, Phys. Rev. D \textbf{89}, 123501 (2014).
%
\bibitem{Koma7}  N. Komatsu, S. Kimura, Phys. Rev. D \textbf{90}, 123516 (2014).
\bibitem{Koma8}  N. Komatsu, S. Kimura, Phys. Rev. D \textbf{92}, 043507 (2015).
\bibitem{Koma9}  N. Komatsu, S. Kimura, Phys. Rev. D \textbf{93}, 043530 (2016).
%






\bibitem{Hooft-Bousso}
G. 't Hooft, arXiv:gr-qc/9310026; L. Susskind, J. Math. Phys. \textbf{36}, 6377 (1995); R. Bousso, Rev. Mod. Phys. \textbf{74}, 825 (2002).









\bibitem{Bekenstein1Hawking1}  
J. D. Bekenstein, Phys. Rev. D \textbf{7}, 2333 (1973); Phys. Rev. D \textbf{9}, 3292 (1974);  Phys. Rev. D \textbf{12}, 3077 (1975); 
S. W. Hawking, Phys. Rev. Lett. \textbf{26}, 1344 (1971); Nature \textbf{248}, 30 (1974);
Phys. Rev. D \textbf{13}, 191 (1976).
%




\bibitem{Tsallis2012}  C. Tsallis, L. J. L. Cirto, Eur. Phys. J. C \textbf{73}, 2487 (2013). 

\bibitem{Czinner1}       T. S. Bir\'{o}, V. G. Czinner, Phys. Lett. B \textbf{726}, 861 (2013).
\bibitem{Czinner2-3}    V. G. Czinner, H. Iguchi, Phys. Lett. B \textbf{752}, 306 (2016);
V. G. Czinner, H. Iguchi, Eur. Phys. J. C \textbf{77}, 892 (2017).




%
\bibitem{LQG2004_1} A. Chatterjee, P. Majumdar, Phys. Rev. Lett. \textbf{92}, 141301 (2004).
\bibitem{LQG2004_2} A. Ghosh, P. Mitra, Phys. Rev. D \textbf{71}, 027502, (2005).
\bibitem{LQG2004_3} K.A. Meissner, Class. Quantum Grav. \textbf{21}, 5245, (2004).

%
\bibitem{Das2008} S. Das, S. Shankaranarayanan, S. Sur, Phys. Rev. D \textbf{77}, 064013 (2008).
\bibitem{Radicella2010} N. Radicella, D. Pav\'{o}n, Phys. Lett. B \textbf{691}, 121 (2010).






\bibitem{Padma1Verlinde1}  
T. Padmanabhan, Mod. Phys. Lett. A \textbf{25}, 1129 (2010); 
E. Verlinde, J. High Energy Phys. 04 (2011) 029.




\bibitem{Visser}  
M. Visser, J. High Energy Phys. 1110 (2011) 140.




%
\bibitem{Barrow22}  
J. D. Barrow, T. Clifton, Phys. Rev. D \textbf{73}, 103520 (2006).


\bibitem{Farooq2017}  O. Farooq, F. R. Madiyar, S. Crandall, B. Ratra, Astrophys. J. \textbf{835}, 26 (2017).



%
\bibitem{Ren1}    A. R\'{e}nyi, \textit{Probability Theory} (North-Holland, Amsterdam, 1970).
%
\bibitem{Tsa0}    C. Tsallis, J. Stat. Phys. \textbf{52},  479 (1988).






\bibitem{Padma2005}  T. Padmanabhan, Class. Quantum Grav. \textbf{22}, L107, (2005).











\end{thebibliography}
\end{document}